# Ensemble Machine Learning Model for Inner Speech Recognition: A Subject-Specific Investigation

Shahamat Mustavi Tasin[1], Muhammad E. H. Chowdhury[2,*], Shona Pedersen[3], Malek Chabbouh[2], Diala Bushnaq[2], Raghad Aljindi[2], Saidul Kabir[1], Anwarul Hasan[4]

[1]Department of Electrical and Electronic Engineering, University of Dhaka, Dhaka-1000. Email: tasin.mustavi@gmail.com (SMT), kabirsaidul116@gmail.com (SK)

[2]Department of Electrical Engineering, Qatar University, Doha 2713. Email: mchowdhury@qu.edu.qa (MEHC); mc1912750@qu.edu.qa (MC); db1901951@qu.edu.qa (DB); aghad.aljindi@qu.edu.qa (RA)

[3]Department of Basic Medical Sciences, College of Medicine, Qatar University, Doha, 2713, Qatar. Email: spedersen@qu.edu.qa (SP)

[4]Department of Industrial and Mechanical Engineering, Qatar University, Doha-2713, Qatar. Email: ahasan@qu.edu.qa (AH)

*Corresponding author: Muhammad E. H. Chowdhury (mchowdhury@qu.edu.qa)

## Abstract

Inner speech recognition has gained enormous interest in recent years due to its applications in rehabilitation, developing assistive technology, and cognitive assessment. However, since language and speech productions are a complex process, for which identifying speech components has remained a challenging task. Different approaches were taken previously to reach this goal, but new approaches remain to be explored. Also, a subject-oriented analysis is necessary to understand the underlying brain dynamics during inner speech production, which can bring novel methods to neurological research. A publicly available dataset, "Thinking Out Loud Dataset", has been used to develop a Machine Learning (ML)-based technique to classify inner speech using 128-channel surface EEG signals. The dataset is collected on a Spanish cohort of ten subjects while uttering four words ("Arriba", "Abajo", "Derecha", and "Izquierda") by each participant. Statistical methods were employed to detect and remove motion artifacts from the Electroencephalography (EEG) signals. A large number (191 per channel) of time-, frequency- and time-frequency-domain features were extracted. Eight feature selection algorithms are explored, and the best feature selection technique is selected for subsequent evaluations. The performance of six ML algorithms is evaluated, and an ensemble model is proposed. Deep Learning (DL) models are also explored, and the results are compared with the classical ML approach. The proposed ensemble model, by stacking the five best logistic regression models, generated an overall accuracy of 81.13% and an F1 score of 81.12% in the classification of four inner speech words using surface EEG signals. The proposed framework with the proposed ensemble of classical ML models shows promise in the classification of inner speech using surface EEG signals.

## Keywords

Inner Speech, Brain-Computer Interface (BCI), Machine Learning, Classification, Convolutional Neural Network (CNN), Deep Learning, Logistic Regression, Ensemble Model

## 1. Introduction

Language and speech processing has always been an intriguing and complicated topic in neurological research. This is because a number of cerebral structures and networks associated with factors like emotional processing, contextual meaning, and cognitive regulation collectively add to the core language system [1]. A number of models have been proposed that underline the necessity of dorsal connection to explain the neuroanatomy of language in a simplified manner [2], while other investigations have provided evidence showcasing the intricate composition of the arcuate fasciculus alongside the superior longitudinal fasciculus [3], [4], [5]. This neural pathway establishes connections between numerous target regions within the frontal cortex and multiple target regions within the temporal and parietal cortex. Moreover, investigations utilizing functional imaging techniques to study healthy subjects predominantly reveal bilateral activations and the involvement of a large network that exhibits partial activity depending on specific language tasks [6]. This finding contrasts with classical models, such as the Broca or Wernicke models, which primarily emphasize the study of the left hemisphere. Interestingly, motor system and sensorimotor pathways have also been found to be involved in language and speech processing [7], [8]. Different speech components such as phonetic properties, semantic content, and even abstract linguistic constructs related to the articulated or inner speech appear to be grounded in the motor system [9], [8], [10]. This connection gives rise to numerous interactions between the motor system and the complex processes of language perception and speech production.

Electrophysiological signals recorded from the brain have been an integral part of research on brain dynamics for decades. Such research has helped scientists understand how the brain processes language and various speech forms. Electrocorticography (ECoG) or intracranial Electroencephalography (iEEG), Stereo Electroencephalography (SEEG), and Electroencephalography (EEG) record brain activity; however, surface EEG has long been the signal of choice due to its non-invasive nature and practicality. Typically, EEG signals are acquired by placing electrodes on the scalp and capturing them with EEG recording systems [11]. EEG signals are valuable inputs for a variety of applications, including emotion detection [12], rehabilitation [13], and, most importantly, speech recognition [14]. Numerous studies have been conducted to classify brain activities associated with various components of language and speech production, including syllables, phonemes, vocals, and words. Zhao and Rudzicz [15] studied seven phonetic and syllabic cues during covert speech production and categorized them. In more recent works, Jahangiri et al. [16], [17] analyzed four phonemic structures. Other studies, including Cooney et al. [18], Tamm et al. [19], and Ghane and Hossain [20], analyzed EEG signals generated during the imagined speech production of five vowels. However, studies focusing on minute components such as vowels, phonemes, and syllables are often limited to a narrow scope of applicability, necessitating additional research on a broader aspect of language and speech. In this regard, there have been other studies that were carried out involving imagined words. González-Castaeda et al. [14] categorized the five imagined words: "up," "down," "left," "right," and "select." Although results were analyzed subjectively, there were no subject-specific investigations into inner speech production. The study by Pawar and Dhage [21] addressed the same prompts with the exception of the word "select". Here, the investigation was centered around specific brain regions, and the authors attempted to devise a generalized solution. Furthermore, Nguyen et al. [22] analyzed two distinct groups of imagined words (short and long). The first group consisted of the words "in", "out", and "up", while the second group was composed of the words "cooperate" and "independent." In this work, channel selection was done based on the investigations into 2-3 subjects out of 15. Thus, it is evident that a subject-specific investigation into EEG signals to decode language and speech perception and production has not been explored enough in the literature. This calls for more research using such an approach to decode brain signals for classifying inner speech. The lack of publicly available EEG dataset to classify words from inner speech makes this a challenging task.

In the current study, we have therefore performed a subject-wise analysis of the EEG signals to distinguish the nature of electrophysiological signals generated when an individual is producing words internally. A dataset prepared by recording EEG signals during inner speech production in Spanish language has been utilized for this matter. Motion artifacts have been removed before extracting features from the raw data. Then, feature-selection has been carried out and inner speech is classified using Machine Learning (ML) techniques for each subject. The main contributions of this study are as follows:

- Motion artifacts were detected using statistical methods and were removed by discarding the low-frequency components using variational mode decomposition (VMD) on the channels exhibiting such artifacts as a pre-processing step.
- 2D visualization techniques such as topographical maps were utilized to identify differences between different states like resting, visual processing, and inner speech production. This was also employed to understand the differences among the subjects in producing inner speech as well as to observe the trends for different words.
- Several time-, frequency-, and time-frequency-domain features were extracted from the EEG signals, followed by utilizing various feature selection algorithms to enhance the performance of ML models.
- Several classical ML-based techniques, 1D signal and 2D spectrogram-based deep learning (DL) techniques were evaluated using the processed as well as the raw data, followed by proposing an ensemble model to classify the inner speech.
- A publicly available dataset for inner speech has been utilized to evaluate the proposed methodology, and the results obtained from our approach surpassed the existing works on this dataset by a significant margin.

Section 3 outlines the results and discussion obtained from all the methods described in the previous section through quantitative and qualitative analysis. Section 4 presents brief conclusions from this study.

## 2. Materials and Methods

For the identification of inner speech from EEG signals, a publicly available dataset [23] has been used. This section provides a detailed discussion of the study materials and methods, which contains dataset description, data processing steps taken by the authors of the dataset, Event-Related Potential (ERP) analysis of the EEG data, motion artifact removal process, feature extraction steps, feature selection methods, and the experiments using different ML and DL models. The overall proposed framework is shown in Figure 1.

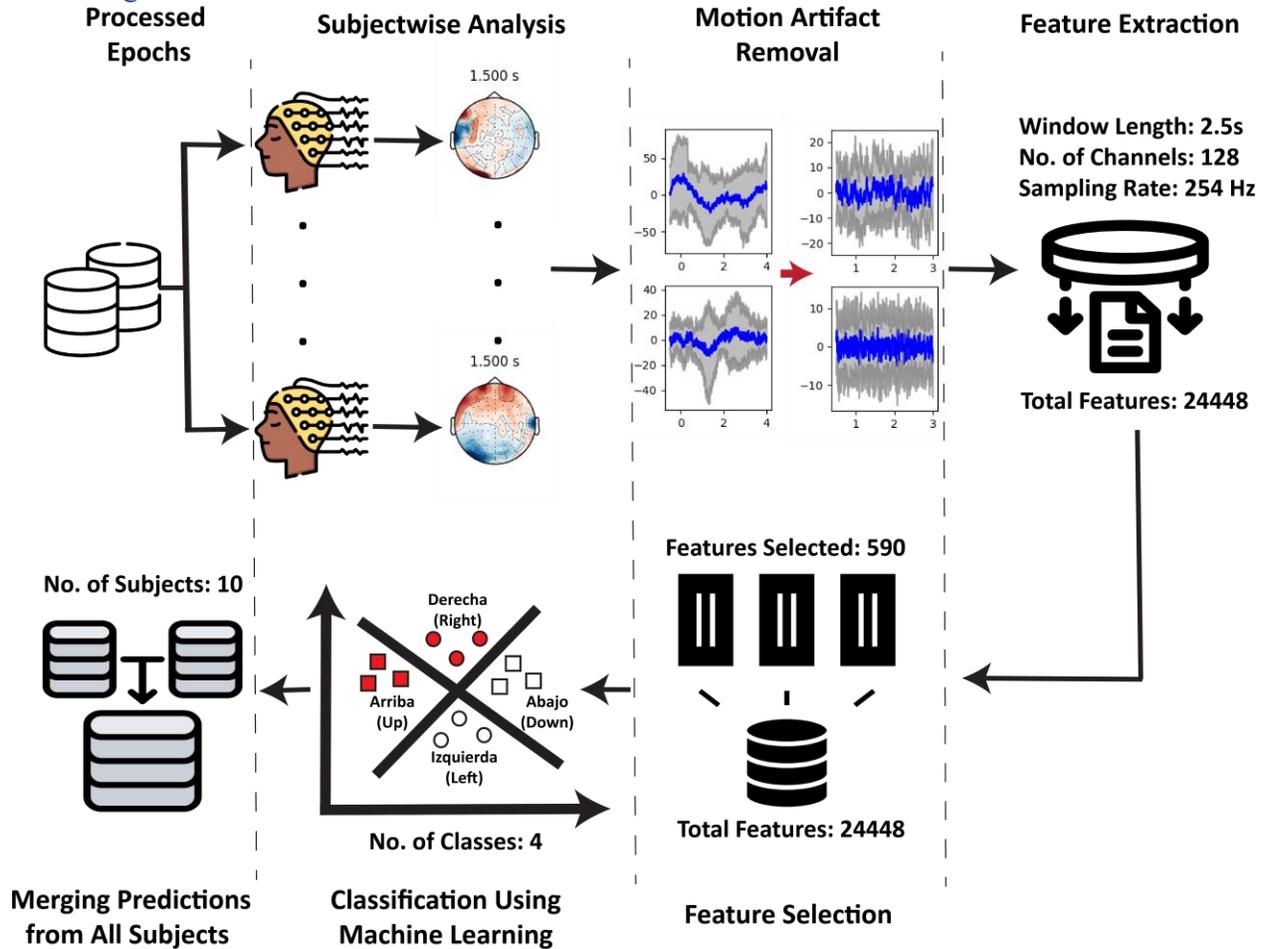

**Figure 1.** Proposed methodology for inner speech detection.

*2.1. "Thinking Out Loud Dataset" Description*

A publicly available dataset shared by Nieto et al. [23] has been used in this study to evaluate the proposed approach. The dataset was collected from the Spanish population while the ten participating subjects were uttering four words ("Arriba", "Abajo", "Derecha", and "Izquierda"). These four words correspond to "Up", "Down", "Right", and "Left", respectively. Even though the dataset aimed to improve understanding of inner speech, the dataset contributors also collected EEG signals while the participants performed pronounced speech and visualized or imagined speech activities. 128 channels of EEG, as well as 8 channels for Electrooculography (EOG) and Electromyography (EMG) signals, were recorded with a 24 bits resolution and a sampling rate of 1024 Hz using "ActiveTwo system, BioSemi" amplifier [23]. The external channels facilitated motor and eye movement detection during inner speech activities to tag erroneous experiments, as well as investigations for pronounced speech detection. Each trial by the participants lasted for 4.5 seconds, of which the corresponding speech activity was performed for 2.5 seconds. A detailed outlook of a sample trial is shown in **Figure *2***. The data collection process involves concentration, cue, action, relaxation, and rest interval. The data collection was conducted in 3 sessions, in which the subjects completed 200 trials in each of the first 2 sessions. However, based on the fatigue and willingness of the subjects to carry out further trials, the number of experiments in the third session varied for different subjects.

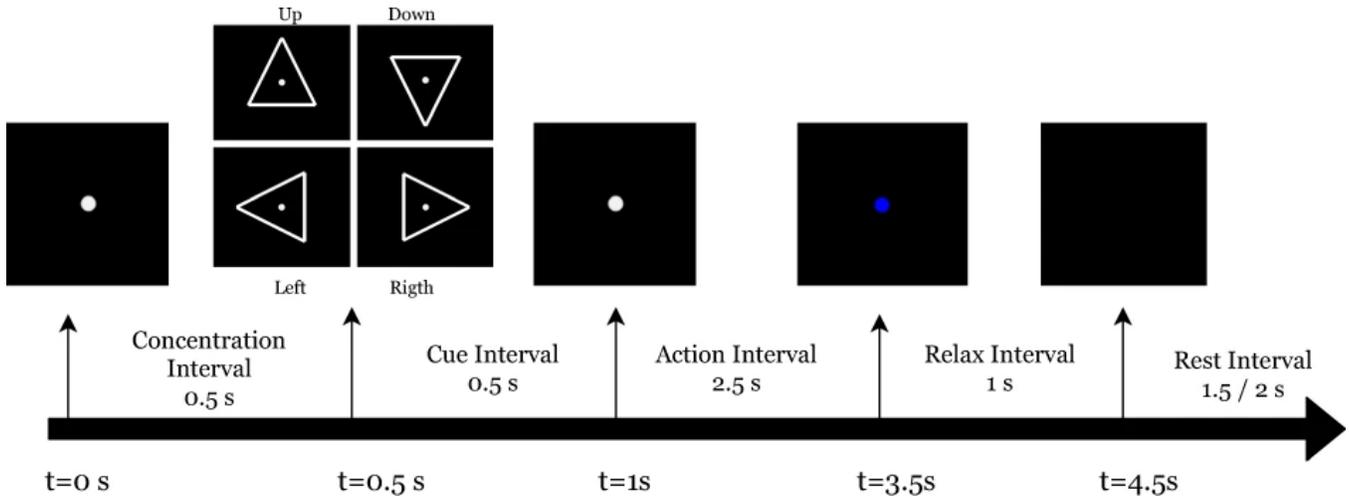

**Figure 2.** The data collection setup. In each trial, one of the four cues ("Up", "Down", "Right", "Left") were shown to the participant [23].

*2.2. Data Processing*

Some processing of the data was done by the authors of this dataset [23]. At first, the data were re-referenced using two external channels as the data collection was done using a reference-free acquisition system. Then, a zero-phase bandpass filter with finite impulse response was applied to the data, the lower and upper bounds being 0.5 Hz and 100 Hz, respectively, followed by applying a notch filter of 50 Hz. After that, the data were decimated to obtain a sampling rate of 256 Hz before segmenting the data. Then, Independent Component Analysis (ICA) was employed on the EEG channels to capture sources, and it aided in detecting unwanted motions, which were discarded during the reconstruction of EEG signals to generate the final dataset. Also, EMG signals were used to detect mouth movements during inner and visualized condition actions, which were used for tagging the trials containing such movements. The raw data, as well as the processed data, are provided in the publicly available dataset [23]; however, the processed data is used for this study.

*2.3. Topography Analysis of EEG Data*

Event Related Potentials (ERPs) are measured using 128 channels EEG amplifier and are used to investigate the behaviour of different brain regions. ERPs are time-locked to specific events or stimuli, such as visual, auditory, or cognitive stimuli. They represent the brain's electrical activity in response to these events and provide insights into the neural processes underlying perception, attention, memory, and other cognitive functions [24]. From ERPs, topographical maps can be constructed, which can provide insights into the activities of different regions during any state.

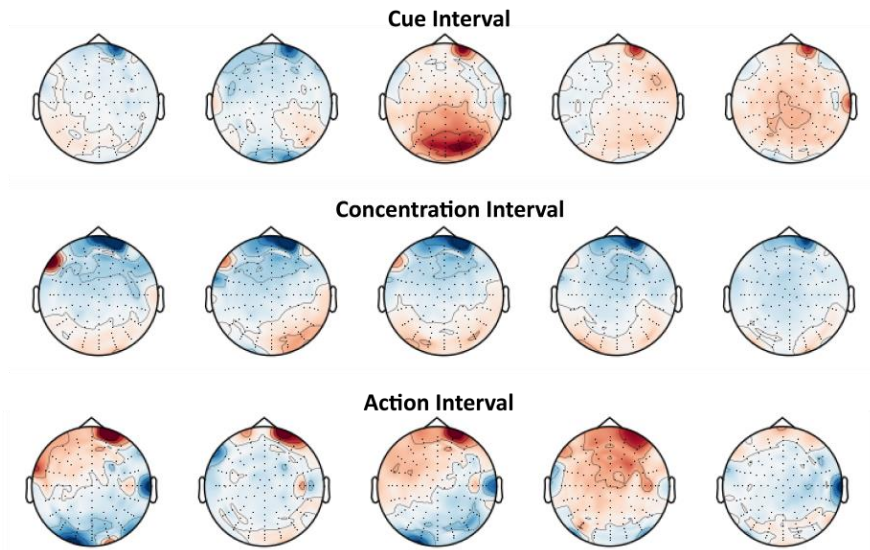

**Figure 3.** Topographical maps of cue interval, concentration interval, and action interval for subject 10.

Analyzing the topographies of different subjects and for different intervals in this dataset reveals that the activities of different regions of the brain were different for all intervals, with more activity in the visual cortex during the cue interval (when participants were shown visual information about the four cues) and intense activities in specific locations during the concentration and action interval, when participants were engaged in focused attention and action, as evident in Figure 3. In addition, it is observed that the ERPs of distinct words uttered internally by a particular subject shared some similarities, although these similarities were weak for the majority of subjects. However, there is no similarity among the topographies of different subjects. Figure 4 suggests that inner speech production of each subject was unique. This necessitates a subject-specific data investigation and evaluation of various approaches.

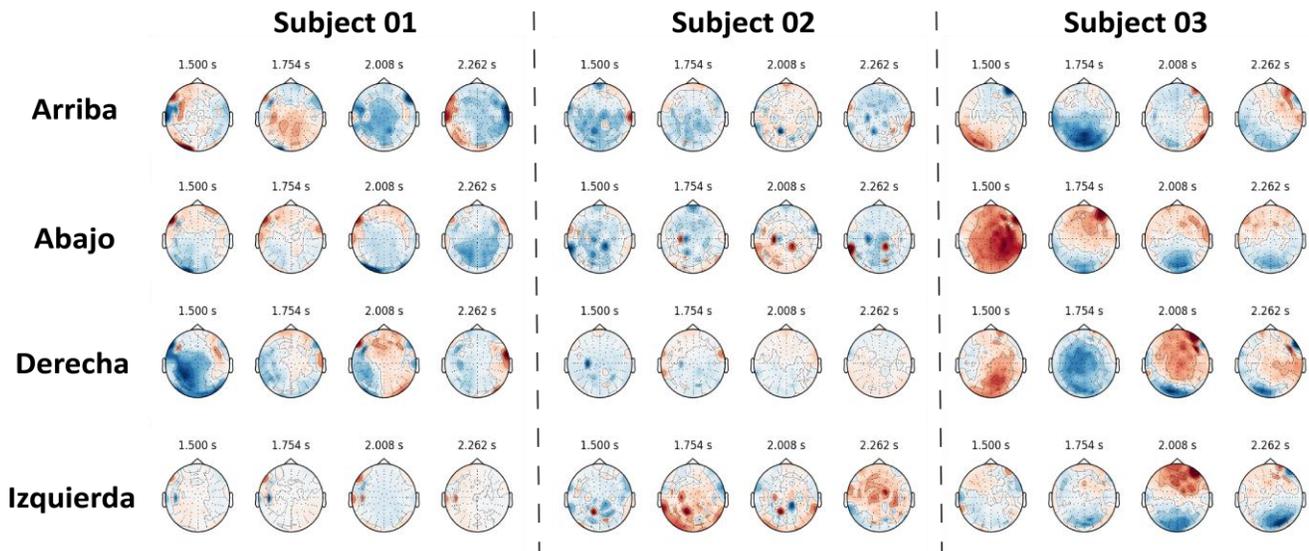

**Figure 4**. Topographical maps of three subjects with all four words uttered during inner speech tasks.

*2.4. Motion Artifact Removal*

Even though some measures were taken to eliminate artifacts from the raw data [23] by the dataset provider, the processed data may still contain motion artifacts. The presence of motion artifacts can substantially hinder the detection of inner speech; therefore, these artifacts must be eliminated for the accurate identification of words from EEG signals. The channels that exhibit such artifacts are identified in this study by closely examining the signals of all the channels of various subjects using mean and standard deviation plots as shown in Figure 5. These artifacts have to be removed in order to clear up the data. Methods such as Discrete Wavelet Transform (DWT) [25], Wavelet Packet Decomposition (WPD) [26], Empirical Mode Decomposition (EMD) [27], Ensemble Empirical Mode Decomposition (EEMD) [28], Singular Spectrum Analysis (SSA) [29], and Variational Mode Decomposition (VMD) [30] were explored previously to remove motion artifacts [31] [32]. VMD is utilized in this study to remove motion artifacts due to its resistance to sensitivity and noise, as well as its non-recursive method for estimating modes concurrently [32]. Using VMD, six modes are generated from the data of each trial, and the Intrinsic Mode Function (IMF) corresponding to the mode with the lowest frequency is discarded [33]. Figure 5 depicts the entire motion artifact removal procedure. After that, the clean data are used to reconstruct topographical maps, and it is observed that ERP drifts have been removed (Figure 6).

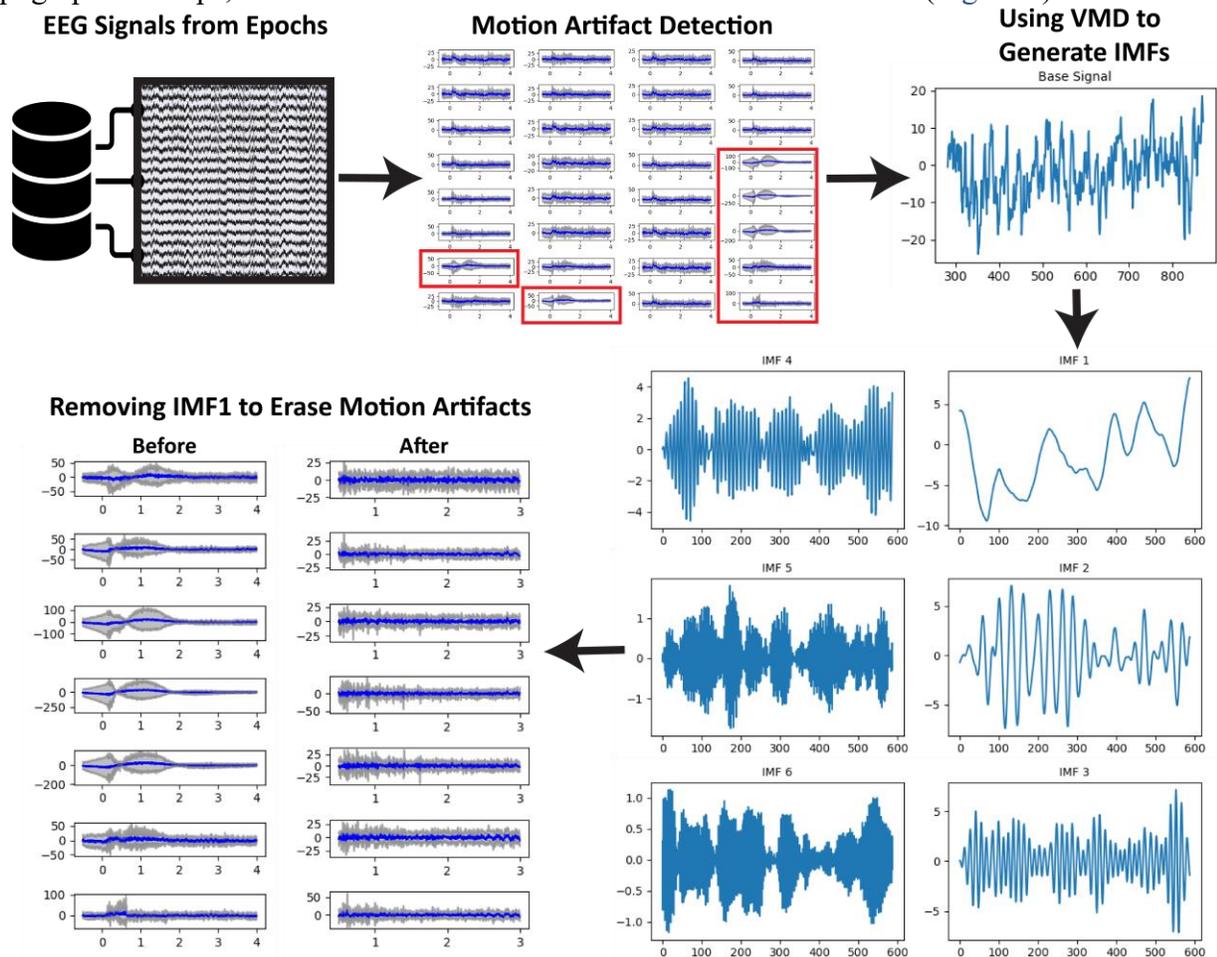

**Figure 5**. Motion artifact detection using statistical methods and removal of artifacts using VMD.

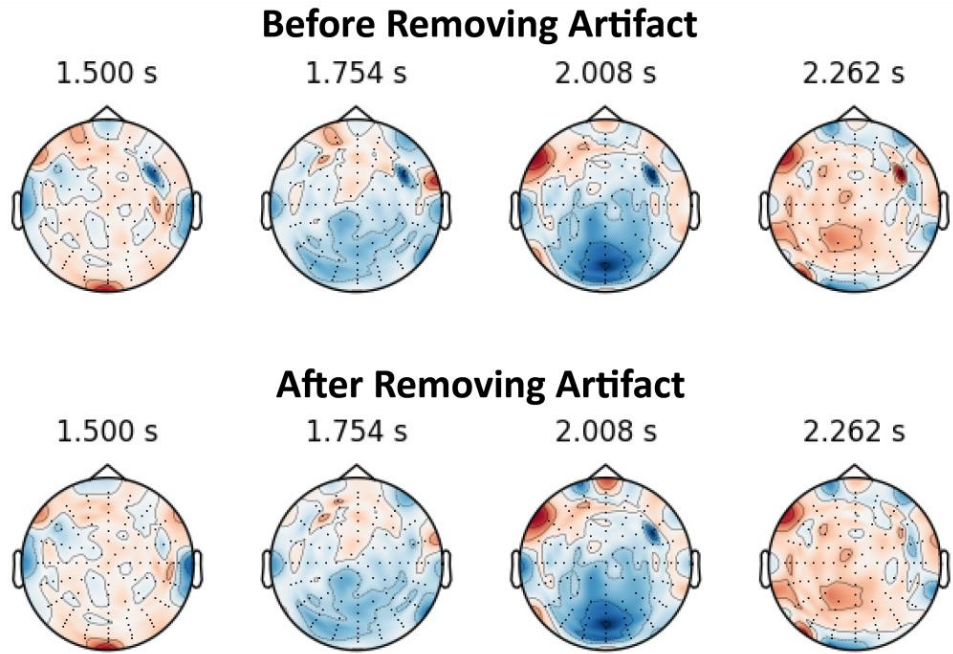

**Figure 6**. Effect of removing motion artifacts (subject 06, class - "Izquierda").

*2.5. Feature Extraction*

Time-domain (TD) [34], frequency domain (FD) [35], [36], and time-frequency domain (TFD) [37] features have been extracted from the EEG signals of the action interval of each subject. The TD features consist of pure TD characteristics and other statistical features (such as mean, deviation, absolute value, etc.) obtained from the signal in the time domain [38]. Features obtained from the envelope of the signal have also been included under TD features. Features that describe the shape and size of the signal in the frequency domain representation have come under the FD features. These features consist of both simple features, such as dominant frequency, and more complex features, such as spectral roll-off point and spectral deformation. Some other FD features have also been calculated, such as skewness and kurtosis of the power spectrum. The TFD features in this study are mainly obtained using discrete wavelet transform (DWT) [39], [40]. The features consist of calculating the band power, mean absolute value, waveform length, root mean square value, standard deviation, and fractal length of the extracted wavelets. In total, 50 TD, 24 FD, and 117 TFD features have been extracted from each EEG channel segment in this study. The total number of features per channel was 191. The details of all the features from different domains are provided in the **Supplementary Tables T1, T2, and T3**. The dataset contains data from 128 EEG channels. As a result, a total of 24448 features (191 × 128) were extracted from the data of each subject. Such a huge number of features can cause a high variance in the prediction using ML models [41] and hence requires feature selection techniques.

*2.6. Feature Selection Techniques*

Following feature extraction, selecting features has become crucial since the huge number of features is likely to cause a significant downgrade of model performance due to overfitting. With that in mind, the following feature selection techniques are explored:
1. ANOVA F [42]
2. Chi-Square [43]
3. Mutual Information (MI) [44]
4. Pearson-Coefficient (PC) [45]
5. Decision Tree-Based Selection [46]
6. Principal Component Analysis (PCA) [47]

7. ReliefF [48]
8. Minimum Redundancy Maximum Relevance (MRMR) [49]

Preliminary experiments showed that the MRMR method is the most suitable approach in this study, and therefore, it is discussed briefly in the following section.

*2.6.1  MRMR*

Minimum Redundancy Maximum Relevance (MRMR) is an algorithm for identifying a subset of relevant features while minimizing redundancy. Using measures such as mutual information and correlation coefficients, it quantifies the relevance of each feature to the objective variable. The feature importance score is calculated using the following equation:

$$score_i = \frac{relevance(f|target)}{redundancy(f|features\ selected\ until\ i-1)} \quad (1)$$

The algorithm selects features with high relevance and low redundancy by evaluating their individual association with the target variable and taking redundancy with previously selected features into account. However, this method aims to select a group of features that work best together, taking the idea of "selecting K best features instead of best K features" [50]. Each iteration adds the feature with the highest relevance and lowest redundancy until the desired number of features is reached. Relevance and redundancy are utilized in Equation (1) to calculate the importance score. The MRMR algorithm strikes a balance between choosing features that are predictive of the target variable and ensuring that they are not extremely correlated [51]. In this work, MRMR is applied to the data of each subject separately in the subject-specific investigation.

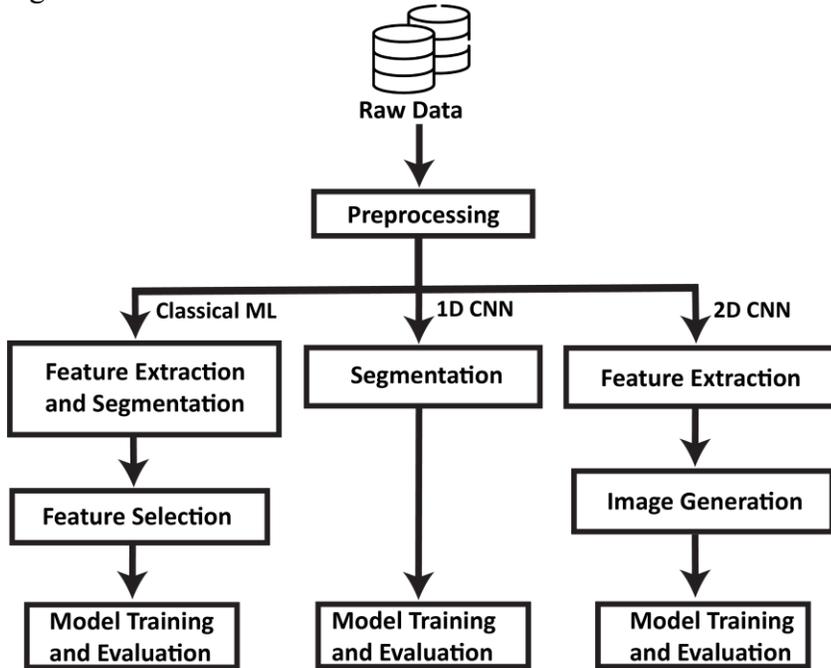

**Figure 7**. Basic methodology of ML and DL techniques.

*2.7. Classification Algorithms*

Each word in the inner speech is considered as a class in this work, making it a four-class classification problem. Six Machine Learning (ML) algorithms, including Support Vector Machine (SVM), Linear Discriminant Analysis (LDA), Extra Trees Classifier (ETC), Random Forest Classifier (RFC), Logistic Regression (LR), and LightGBM (LGBM) are evaluated using 10-fold Cross-Validation (CV) after feature extraction and selection. It should be noted that a subject-specific strategy is used to select features. Evaluation of models is also conducted in this manner, as our investigation revealed that the characteristics of topographies during inner speech production are subject-specific. Previously, the ensemble of ML models proved effective for binary-class and multiclass classification problems [52] [53]. Such approach, along with model stacking, is also observed in classification problems with EEG signals [54]. In light of this, the best models from our evaluation are combined to create ensemble models, and their performance is then evaluated.

DL methods have also been explored by evaluating different one-dimensional (1D) Convolutional Neural Network CNN and two-dimensional (2D) CNN architectures. For 1D CNN approach, a two-layered CNN architecture has been used (see **Supplementary Figure S3** for more details). Each layer comprises of a convolution layer with kernel size set to "3", a maxpooling layer of kernel size 4, and a dropout layer of 10% probability. A dropout layer of 50% probability has also been added to the Multi-Layer-Perceptron (MLP) layer at the output. Segmented EEG time series data from each subject has been fed to the 1D CNN model and the results are recorded.

For the evaluation of 2D CNN approaches, images are generated from the extracted features where each trial corresponds to an image by being constructed from the 2D matrix (n_features × n_channels), as shown in Figure 8. The images have been augmented with rotations (90°, 180°, and 270°) and image translations consisting of 10% vertical-up shift and horizontal-right shift, and 10% diagonal shifts toward top-right and bottom left respectively. The combined set of raw and augmented images are used to classify different words, and the results from two 2D CNN architectures are recorded. "Resnet18" [55] and "VGG16" [56] models have been utilized to evaluate the 2D CNN approach with a 5-fold Cross-Validation (CV) scheme. It it to be noted that, since a subjectwise evaluation has been followed, the size of the training data is small in each case, for which architectures with fewer parameters have been chosen in both 1D and 2D CNN methods to prevent overfitting. The basic methodology of the ML and DL techniques are illustrated in Figure 7.

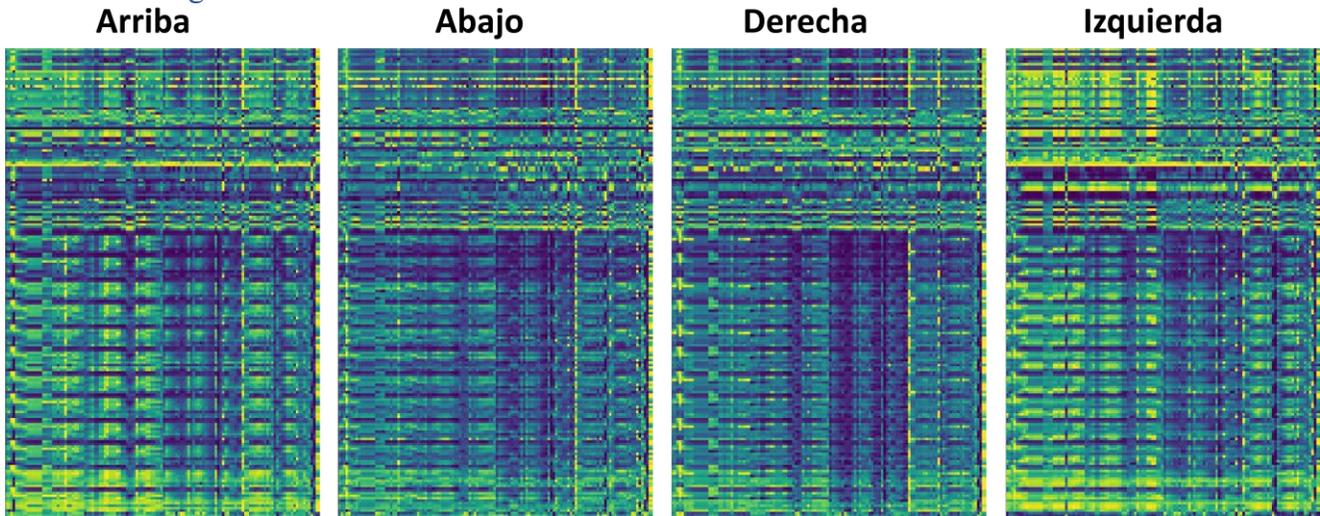

**Figure 8.** Sample images used for 2D CNN approach (subject 01).

*2.8. Evaluation Criteria*

Two metrics, accuracy and F1 score are used to evaluate the performance of the classifiers. True Positive (TP), True Negative (TN), False Positive (FP), and False Negative (FN) values are utilized to generate these scores. It is to be noted that since the dataset is completely balanced for all subjects, accuracy and F1 score are very close and either of these scores can represent the predictive performance of a classification model in this case. The metrics used in this work are discussed below:

1) Accuracy: The accuracy metric measures the proportion of true predictions made by a classifier. It expresses the accuracy as the ratio of accurate predictions to the total number of predictions. As demonstrated by Equation (2), accuracy is determined by dividing the sum of True Positives and True

Negatives, which represents total correct predictions, by the sum of True Positives, True Negatives, False Positives, and False Negatives, which represents all of the predictions.

$$Accuracy\ (\%) = \frac{TP + TN}{TP + TN + FP + FN} \times 100\% \qquad (2)$$

2) F1 Score: F1 score measures the classification accuracy of a model by considering both the correct positive predictions (precision) and the ability to capture true positive instances (recall). The F1 score is calculated as the harmonic mean of precision and recall, providing a well-rounded evaluation of a model's performance. In this instance, the F1 score is calculated using Equation (3), using True Positives, False Positives, and False Negatives.

$$F1\ Score\ (\%) = \frac{2 \times TP}{2 \times TP + FP + FN} \times 100\% \qquad (3)$$

For evaluation of ML classifers and 1D CNN model, a 10-fold CV scheme has been followed. In this method, the data is split into ten folds where nine folds are used for model training and the remaining fold is used for validation and this process is repeated ten times. In each iteration, the predictions by the classifier on the validation set are stored. For each subject, at the end of all iterations, the combined predictions from all ten folds are compared against the ground truths, and accuracy and F1 score are calculated using the Equations (2) and (3) respectively. The overall scores are generated by taking the average accuracy and F1 score of all the subjects, and are used to validate the efficacy of the feature selection methods as well as the performance of all the ML and DL classifiers. The 5-fold CV scheme used in 2D CNN approach follows the same procedure as the 10-fold CV scheme, differing only by the number of folds for each subject.

## 3. Results and Discussion

This study evaluates the efficacy of eight feature selection techniques, with one technique (MRMR) standing out and being selected for all subjects. Analysing the effect of various parameters on particular subjects leads to the selection of optimal hyperparameters in feature selection. Results from classical Machine Learning (ML) algorithms, 1D CNN, and 2D CNN algorithms are presented, demonstrating the efficacy of ensemble ML models. Existing works are outperformed by the proposed ensemble model, while the limited success of deep learning approaches is also discussed. Subject-by-subject results of the ensemble model created through stacking is examined and how the data of the subjects responded better or worse to the proposed ensemble model is explained.

*3.1. Selection of Feature Optimization Technique*

Different feature selection methods, including ANOVA, Chi-square, Mutual Information (MI), Pearson Coefficient (PC), Tree Based Selection using Extra Trees Classifier (ETC), Principal Component Analysis, ReliefF, and MRMR, are evaluated in this work utilizing the six ML models for classification. Table 1 displays the overall results of feature selection and classification for all subjects. Here, ten features are chosen to evaluate the performance of each feature selection method, and only the results from the model with the best overall performance are displayed. With an overall accuracy of 66.53% and an F1 score of 67.09% using an SVM model, it is evident from this analysis that MRMR is the best feature selection method for this data. Other techniques, such as ReliefF and Tree-Based-Selection, also show promise with overall accuracies of 58.88% and 56.13% respectively. However, MRMR has demonstrated significantly better results by outperforming these two techniques by more than 7.5% in accuracy. Nevertheless, feature-set optimization is required to enhance classification results and produce a more robust ML model. In order to determine the optimal feature-set for all subjects, the number of selected features in the MRMR algorithm has been optimized.

Table 1. Performance of different feature selection techniques

| Feature Selection Method | Model Name | Accuracy (%) | F1 Score (%) |
|---|---|---|---|
| ANOVA | LightGBM (LGBM) | 45.94 | 46.3 |
| Chi-square | Random Forest Classifier (RFC) | 42.31 | 42.36 |
| Mutual Information (MI) | LGBM | 50.85 | 51.09 |
| Pearson Coefficient (PC) | LDA | 48.94 | 49.94 |
| Tree Based Selection using Extra Trees Classifier (ETC) | ETC | 56.13 | 56.47 |
| PCA | LGBM | 38.52 | 38.49 |
| ReliefF | ETC | 58.88 | 59.29 |
| **MRMR** | **SVM** | **66.53** | **67.09** |

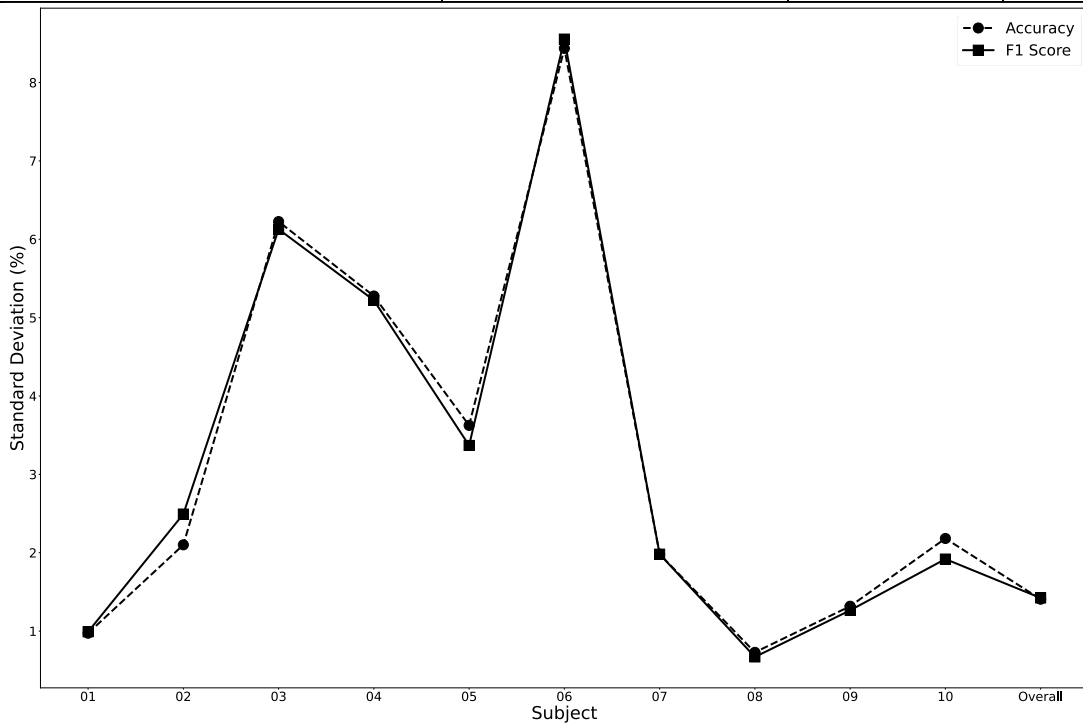

**Figure 9.** Subject-wise variation of results at different feature-sets.

It is observed that the results of subject 06 are impacted the most by the number of features selected by the MRMR algorithm (Figure 10), with a Relative Standard Deviation (RSD) of over 8% for both accuracy and F1 score. High deviations are also observed in the case of subjects 03 and 04 with 6.2% and 5.2% RSD of accuracy respectively. However, further analysis reveals that while there has been a steady increase in the results of subject 06 as the feature-set is increased (Figure 10), the other two subjects show fluctuations in the results for different feature-sets. As the accuracy of subject 06 has increased by 20.84% from its initial state, subjects 03 and 04 have shown 7.22% and 1.25% decrease in accuracy which is minimal compared to the significant improvement of the results of subject 06. The overall results have shown to be unaffected by this hyperparameter tuning, with an RSD of only 1.4%.

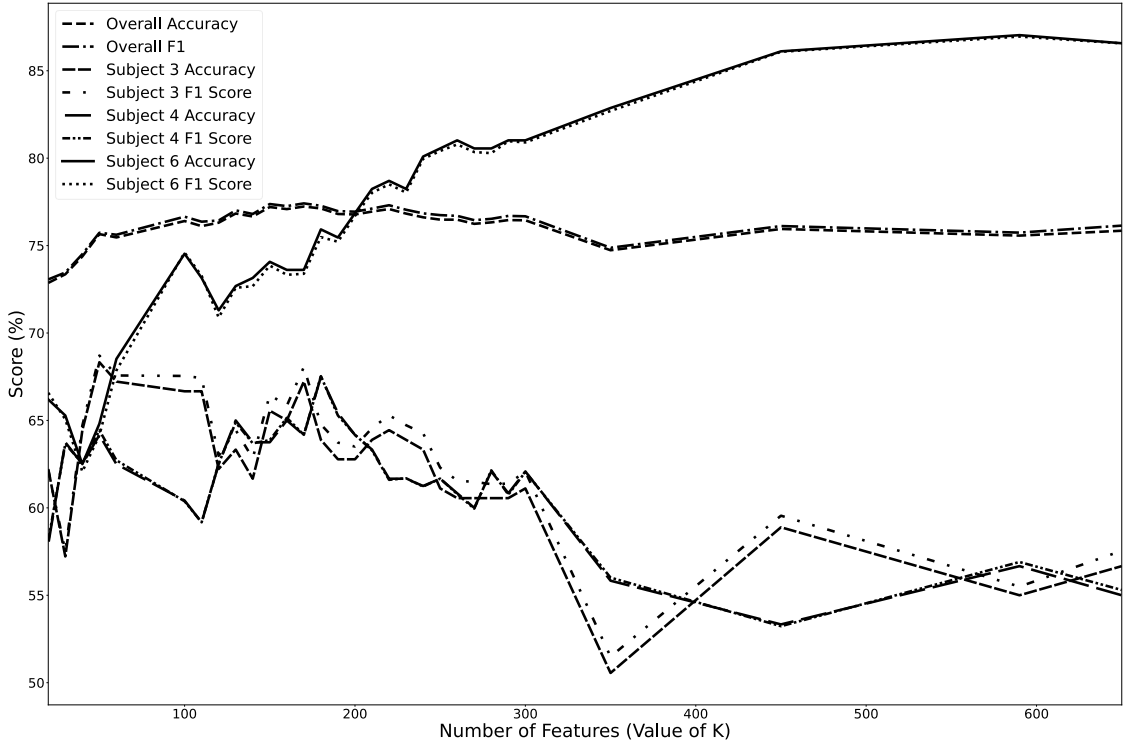
**Figure 10.** Results for hyperparameter tuning of MRMR feature selection.

In MRMR algorithm, the number of features is characterized by the parameter "K" and it is varied from 20 to 610 in this analysis. From Figure 10, subject 06 shows an accuracy and F1 score of 87.04% and 86.95% respectively at K = 590, which is the highest among all the feature sets. As a result, 590 features are selected using the MRMR method for each subject, and the performance of different ML and DL models are carried out according to the evaluation methods described in Section 2.7 and 2.8.

*3.2. Classification Results*

Following feature optimization, different classical ML algorithms, including Support Vector Machine (SVM), Linear Discriminant Analysis (LDA), Extra Trees Classifier (ETC), Random Forest Classifier (RFC), Logistic Regression (LR), and LightGBM (LGBM) are evaluated, with their overall results recorded. It is observed from Table 2 that the LR model has produced 77.89% accuracy and 77.91% F1 score in this dataset, which is the highest among the basic ML models. SVM and ETC models have also shown promising results with 76.31% and 75.93% accuracy respectively, suggesting the evaluation of an ensemble model constituting these three models. Accordingly, the performance of an ensemble model, comprising of one SVM, one ETC, and one LR model is found to be better than that of the LR model, with a 1.22% and 1.18% increase in accuracy and F1 score respectively. However, another ensemble model comprising of five LR models is also constructed and its performance is found to surpass the previous ensemble model, with 3.24% increase in overall accuracy and 3.21% increase in overall F1 score from the LR model. Due to the robust results and less complexity, this ensemble LR model is proposed in this approach.

**Table 2.** Performance of ML and DL approaches, as well as results from the literature

| Model Name | Overall Accuracy (%) | Overall F1 Score (%) |
|---|---|---|
| Support Vector Machine (SVM) | 76.31 | 76.4 |
| Linear Discriminant Analysis (LDA) | 71.91 | 71.89 |
| Extra Trees Classifier (ETC) | 75.93 | 76.16 |
| Random Forest Classifier (RFC) | 73.42 | 73.58 |
| Logistic Regression (LR) | 77.89 | 77.91 |
| LightGBM (LGBM) | 75.33 | 75.43 |
| 1D CNN | 23.732 | 23.173 |
| 2D CNN (Resnet18) | 31.826 | 31.04 |
| 2D CNN (VGG16) | 28.004 | 25.7705 |
| Ensemble of SVM, ETC, and LR models | 79.11 | 79.09 |
| **Ensemble of 5 LR models** | **81.13** | **81.12** |
| EEGNet [57] | 29.67 | 29.61 |

It is noteworthy that the 1D and 2D CNN approaches have produced very poor results compared to the classical ML models. It is because deep learning models often require a lot of data and may fail to reach convergence when the data is not sufficient. Since during training of the models, only data of one subject is provided, it has caused a low number of samples, with number of trials (which equates to the number of samples) only reaching as high as 240. In case of 2D CNN approach, images were augmented with seven augmentations. Likewise, the maximum number of images in the training set at each fold were 304 for each class, which is found to be insufficient in this case. However, when compared with the existing work done on this dataset in which a 2D CNN approach was followed [57], our approach with 2D CNN (Resnet18 architecture) has produced better results with an increase in accuracy and F1 score by 2.15% and 1.43% respectively. Our proposed ensemble model has surpassed the results of the previous work done on this dataset by an astonishing margin and the subject-wise results of this model are recorded in Table 3. The overall consfusion matrix is shown in Figure 11 which depicts the classification efficacy of the proposed model for each class compared to the other classes.

Table 3. Subject-wise results for the ensemble model

| Subject ID | Accuracy (%) | F1 Score (%) |
|---|---|---|
| 01 | 86.50 | 86.38 |
| 02 | 90.83 | 90.82 |
| 03 | 72.22 | 72.16 |
| 04 | 55.00 | 55.09 |
| 05 | 69.17 | 69.17 |
| 06 | 91.67 | 91.62 |
| 07 | 89.17 | 89.19 |
| 08 | 93.00 | 92.97 |
| 09 | 82.92 | 82.88 |
| 10 | 80.83 | 80.88 |

From the subject-wise results shown in Table 3, it is observed that most of the subjects have produced very high classification results with accuracy as high as 93% (subject 08), suggesting that the proposed methodology has the ability to achieve even higher accuracy levels under particular conditions. The consistently high accuracy and F1 scores across the majority of subjects demonstrate the efficacy of the approach. Except three subjects (03, 04, and 05), all subjects have shown accuracy and F1 score of over 80%, causing the overall results to be promising and indicative of the effectiveness of the proposed methodology. The overall confusion matrix supports this conclusion with classification accuracy at near or more than 80% for all the classes.

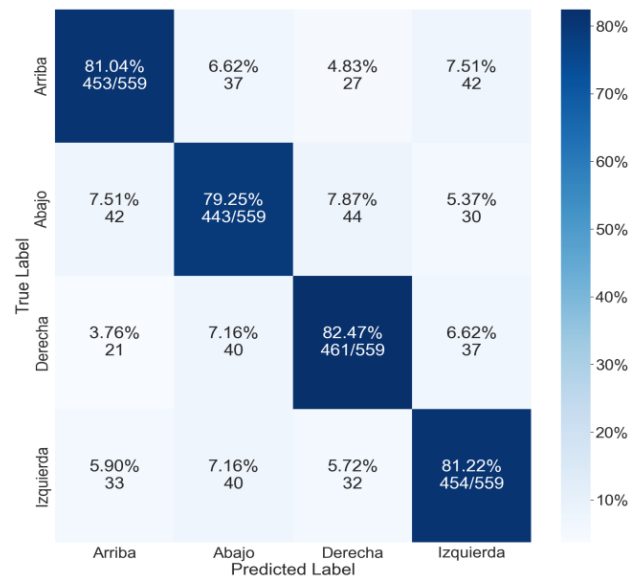

**Figure 11.** Overall confusion matrix for all subjects.

It is noteworthy that the overall performance of the proposed approach has taken a considerable hindrance due to the low performance of the model on the data of subjects 03, 04, and 05. From the confusion matrix of subject 04 (Figure 12(a)), it is observed that the ensemble model could not distinguish well among the classes. This can be explained from Figure 12(b), where it is evident that the same regions of the brain have been active during inner speech production for this subject, which is a strong indicator for confusing one word with another by the model. This trend is also observed in case of subjects 03 and 05 (see **Supplementary Figure S2(c) and (d)** for more details), providing further evidence to this conclusion. It is observed that the words "Arriba" and "Izquierda" are more confused by the model, with classification accuracy being 51.67% for both words. While the predictive performance is low for all classes, it is still well above chance, suggesting that inner speech classification for this subject is possible with a decent accuracy.

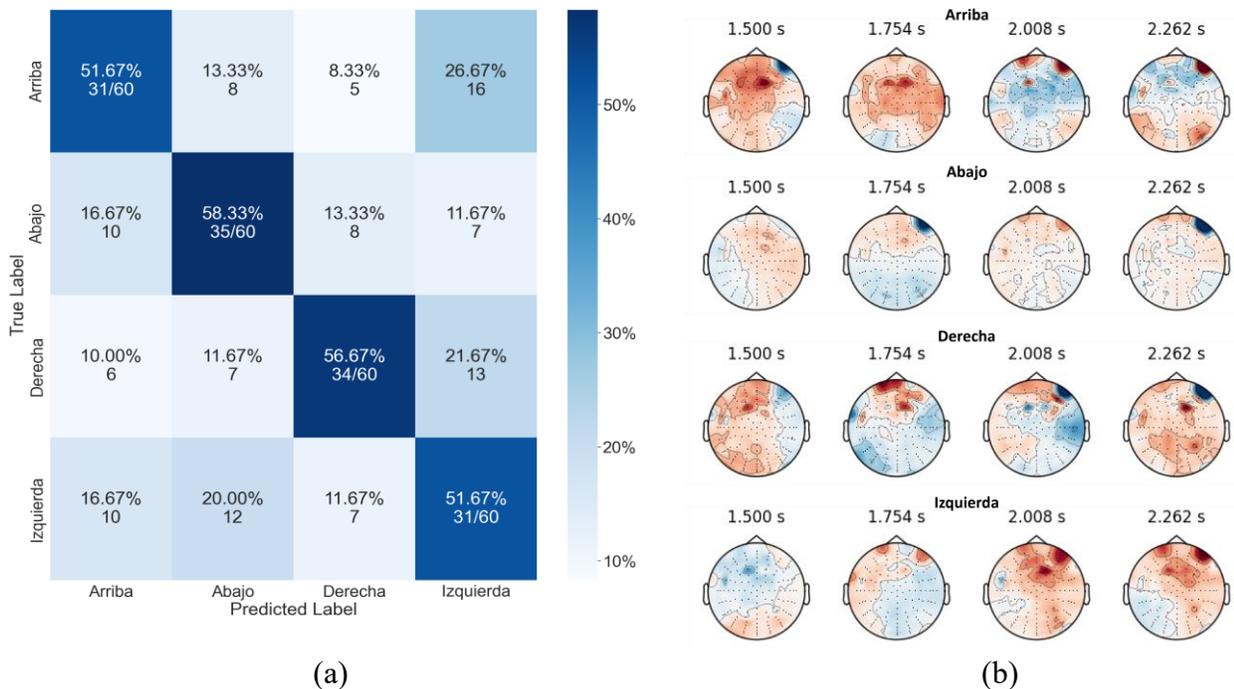

**Figure 12.** (a) Confusion matrix and (b) topographical maps of action interval for subject 04.

However, for other subjects, such as subject 08, different regions are shown to be active during inner speech production for different words (Figure 13(b)), suggesting that the model could differentiate among the words easily. This is supported by the confusion matrix shown in Figure 13(a) which illustrates the strong predictive performance of the model for this subject. It is observed from the confusion matrix that the word "Izquierda" is sometimes confused with other words although the amount of confusion is very little, only as high as 8% with the word "Abajo". This can be explained by the topographical maps depicted in Figure 13(b), showing that only a handful of channels in the occipital region is active during inner speech production of this word, and those chnnels are common for other words as well. Regardless, the proposed ensemble model could accurately distinguish all the words in case of this subject.

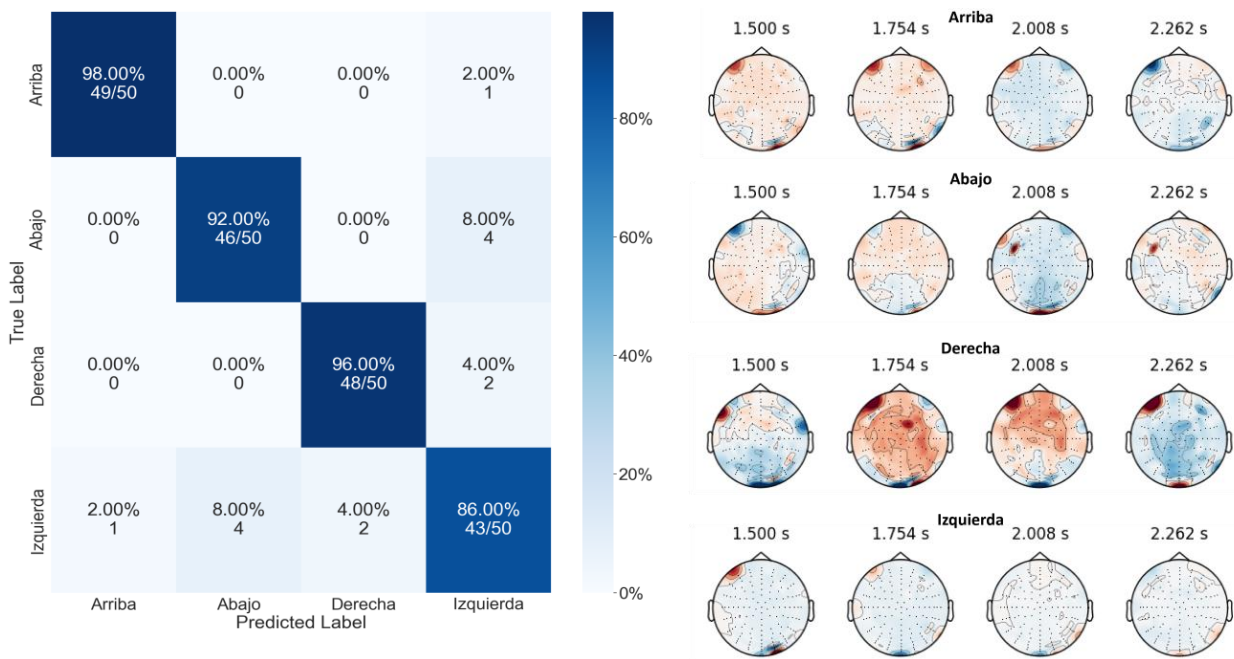

(a) (b)

**Figure 13.** (a) Confusion matrix and (b) topographical map of action interval for subject 08.

The ensemble model constructed by stacking five LR models outperforms all other ML and DL approaches. Feature selection using MRMR is vital to this approach as it can select the most important features out of a large set of 24448 features. The ensemble model, combined with a smaller set of features, reduces model complexity and shows promising results, and demonstrates a possible direction in accurate identification of inner speech.

**4. Conclusion**

In this study, a publicly available dataset has been used for inner speech recognition from EEG signals. Data were collected from 10 healthy subjects, and four words were uttered by each of them during inner speech tasks. The data have been pre-processed and features are extracted in different domains. Eight feature selection techniques were evaluated to select the best features before training on ML models. The performance of the ML models as well as three DL models, were observed. For evaluation, a subject-specific approach was taken due to the unique nature of creating inner speech by each subject. Finally, an ensemble model is proposed to classify inner speech that has produced the best results in this dataset. However, due to the subject-oriented nature of feature selection and model evaluation, the generalizability of this approach is limited. In future, more generalized approaches can be explored where inner speech can be classified with reasonable accuracy by combining data from multiple subjects. The scope of 1D CNN approach can be expanded by applying channel selection methods like Common Spatial Pattern (CSP) algorithm and Gradient-Class Activation Mapping (Grad-CAM). Additional image augmentation techniques can be explored for 2D CNN approach to generate more images, but it may require more time for model training. Overall, the proposed ensemble model, combined with MRMR feature selection technique, proves to be a viable approach to classify inner speech.

**Funding:** This work was made possible by Undergraduate Research Experience Program grant # UREP29-043-3-012 from Qatar National Research Fund (QNRF). The statements made herein are solely the responsibility of the authors.

**Institutional Review Board Statement:** Not applicable

**Informed Consent Statement:** Not applicable

**Conflicts of Interest:** Authors have no conflict of interest to declare.

**Data Availability Statement:** The processed dataset used in this study can be made available upon a reasonable request to the corresponding author.